\documentclass[11pt,onecolumn,draftcls]{IEEEtran}
\usepackage{graphics}
\usepackage{graphicx}
\usepackage{epsfig,amssymb}
\usepackage{amsmath}
\usepackage{times}
\usepackage{mathrsfs}
\usepackage{array}
\usepackage{amsmath, latexsym, amsfonts, amssymb}
\usepackage[mathscr]{eucal}
\usepackage{graphicx}
\usepackage{array, tabularx, colortbl, color, threeparttable}
\usepackage[table]{xcolor}
\usepackage{psfrag}
\usepackage{multirow}
\usepackage{booktabs}
\usepackage{epsfig}
\usepackage{cite}

\newcommand{\paren}[1]{\left(#1\right)}
\newcommand{\sqparen}[1]{\left[#1\right]}
\newcommand{\brparen}[1]{\left\{#1\right\}}
\newcommand{\field}[1]{\ensuremath{\mathbb{#1}}}
\newcommand{\R}{\ensuremath{\field{R}}} 
\newcommand{\Rp}{\ensuremath{\R_+}} 
\newcommand{\Inb}[1]{\ensuremath{\mathbf{1}_{#1}}} 
\newcommand{\PR}[1]{\ensuremath{\mathsf{Pr}\left\{#1\right\}}} 
\newcommand{\PRW}{\ensuremath{\mathsf{Pr}}} 
\newcommand{\EW}{\ensuremath{\mathsf{E}}} 
\newcommand{\sinr}{\ensuremath{{\rm SINR}}}

\newcommand{\BO}[1]{\ensuremath{O\paren{#1}}} 

\renewcommand{\vec}[1]{\ensuremath{\boldsymbol{#1}}} 

\newtheorem{theorem}{Theorem}
\newtheorem{lemma}{Lemma}
\newtheorem{definition}{Definition}

\begin{document}
\title{Vector Broadcast Channels: \\ Optimality of Threshold Feedback Policies}
\author{\authorblockN{Tharaka Samarasinghe, Hazer Inaltekin and Jamie Evans\\}
\authorblockA{Department of Electrical and Electronic Engineering, The University of Melbourne, Australia.\\ Email: \{ts, hazeri, jse\}@unimelb.edu.au}}
\date{}
\bibliographystyle{ieeetr}
\maketitle

\begin{abstract}
Beamforming techniques utilizing only partial channel state information (CSI) has gained popularity over other communication strategies requiring perfect CSI thanks to their lower feedback requirements. The amount of feedback in beamforming based communication systems can be further reduced through selective feedback techniques in which only the users with channels good enough are allowed to feed back by means of a decentralized feedback policy.  In this paper, we prove that thresholding at the receiver is the rate-wise optimal decentralized feedback policy for feedback limited systems with prescribed feedback constraints.  
This result is highly adaptable due to its distribution independent nature, provides an analytical justification for the use of threshold feedback policies in practical systems, and reinforces previous work analyzing threshold feedback policies as a selective feedback technique without proving its optimality.  It is robust to selfish unilateral deviations.  Finally, it reduces the search for rate-wise optimal feedback policies subject to feedback constraints from function spaces to a finite dimensional Euclidean space.
\end{abstract}

\section{Introduction}

Opportunistic beamforming is a low-feedback low-complexity communication technique in which a base station (BS) forms multiple random (orthonormal) beams, and utilizes multiuser diversity by picking the best user with the highest signal-to-interference-plus-noise-ratio ($\sinr$) per beam for communication.  It was first introduced by Viswanath et. al. for a single beam broadcasted from a BS \cite{Tse02}, and then extended to the communication scenario in which multiple beams are broadcasted from a BS by Shariff and Hassibi \cite{hassibi}.  Its key benefit is to operate based on partial CSI, yet to attain data rates near the capacity limits achievable by complex full CSI based signaling techniques \cite{costa, shamai04}.  Its main drawback is growing numbers of users contending for the uplink communication for feedback as the total number of users in the system increases.  Selective feedback techniques \cite{Gesbert04} can be employed to alleviate this burden on the uplink to some extent. This paper focuses on selective feedback techniques, and formally establishes the structure of rate-wise optimal feedback policies for vector broadcast channels under finite feedback constraints.

Threshold feedback policies arise as the most common technique to implement selective feedback techniques 
\cite{Gesbert04, Xu08, Diaz06, Pugh10, Tharaka-ICC}.  According to a threshold feedback policy, a user feeds back if and only if the $\sinr$ value of a beam is above a predetermined threshold value.  
If threshold values are selected correctly, \cite{Diaz06}  and \cite{Tharaka-ICC} showed that average number of users contending for the uplink channel access can be reduced from an $\BO{n}$ growth to a growth of $\BO{\paren{\log n}}$ \cite{Diaz06} and $\BO{\paren{\log n}^\epsilon}$ \cite{Tharaka-ICC}  while maintaining optimal throughput scaling at the downlink, where $n$ is the total number of users in the system and $\epsilon \in (0, 1)$.

In all these previous works, it was implicitly assumed that threshold feedback policies are optimal to maximize communication rates without any formal proof.  This assumption is intuitive (and even holds in the limit without feedback constraints \cite{Diaz06, Tharaka-ICC}), but its proof is not straightforward due to coupling effects of users' feedback rules on the aggregate rate function over multiple beams for finite systems with finite feedback constraints.  In \cite{Tharaka-ICC}, the optimality of threshold feedback policies among the class of homogenous feedback policies was briefly discussed.  In this work, we formally analyze the optimality of more general threshold feedback policies by allowing them to be heterogenous under finite feedback constraints.  Unlike \cite{hassibi, Diaz06, Pugh10} focusing on the asymptotic throughput scaling with feedback amounts (i.e., the number of users contending for the uplink channel access) growing large as the total number of users increases, our results in this paper hold for any finite number of users with finite feedback constraints.


Our contributions and the organization of the paper are as follows.  We precisely define feedback policies and formulate the problem of finding the optimal feedback policy maximizing aggregate communication rate under finite feedback constraints as a function optimization problem in Section \ref{Section: System Model}.  We prove that the rate-wise optimal feedback policy solving this optimization problem is a threshold feedback policy in Section \ref{Section: Optimality of Thresh}.  Section \ref{Section: Discussion} provides a discussion on these results, and Section \ref{Section: Conclusions} concludes the paper.  Our results are distribution independent, provide an analytical justification for the use of threshold feedback policies in practical systems, and strengthen previous work on thresholding as an appropriate selective feedback scheme.  They also form a basis for the optimum threshold selection problem analyzed in our companion paper \cite{Tharaka-ISIT2}. 

\section{System Model} \label{Section: System Model}
Consider a vector broadcast channel in which a BS communicates with mobile users through $M$ different beams simultaneously.  The BS has $M$ transmitter antennas, and users have one receiver antenna.  The beams are assumed to be statistically identical, and users experience statistically independent channel conditions.  We let $\gamma_{i, m}$ be the $\sinr$ at beam $m$ at user $i$ with a continuous distribution.  We also let $\mathcal{N}= \brparen{1,2,\ldots,n}$ and $\mathcal{M}= \brparen{1,2,\ldots,M}$.

Each user calculates $\sinr$ values at all beams, and $\vec{\gamma}_i = \paren{\gamma_{i, 1}, \gamma_{i, 2}, \cdots, \gamma_{i, M}}^\top \in \Rp^M$ represents the $\sinr$ vector at user $i$.  If $M = 1$, we will use $\gamma_i$ to denote the $\sinr$ of user $i$ on this single beam. Let $\gamma_i^\star = \max_{1\leq k \leq M} \gamma_{i, k}$ be the maximum $\sinr$ value at user $i$, 
and let $b_i^\star = \arg\max_{1 \leq k \leq M} \gamma_{i, k}$ be the index of this beam. The $M$-by-$n$ $\sinr$ matrix of the whole $n$-user communication system is denoted by $\vec{\Gamma} = \sqparen{\vec{\gamma}_1, \vec{\gamma}_2, \cdots, \vec{\gamma}_n} \in \Rp^{M \times n}$.  We formally define a feedback policy as follows. 
\begin{definition}\label{fb policy}
A feedback policy $\vec{\mathcal{F}}: \Rp^{M \times n}\mapsto
\brparen{\Omega \bigcup \{ \o \}}^n$ is an $\brparen{\Omega
\bigcup \{ \o \}}^n$-valued function $\vec{\mathcal{F}} =
\paren{\mathcal{F}_1,\mathcal{F}_2,\cdots,\mathcal{F}_n}^\top$,
where $\mathcal{F}_i: \Rp^{M \times n} \mapsto \Omega \bigcup \{
\o \}$ is the feedback rule of user $i$, $\Omega$ is the set of
all feedback packets and $\o$ represents the no-feedback state. We
call $\vec{\mathcal{F}}$ a {\em general decentralized feedback
policy} if $\mathcal{F}_i$ is only a function of $\vec{\gamma}_i$
for all $i \in \mathcal{N}$.  We call it a {\em homogenous general
decentralized feedback policy} if all users use the same feedback
rule, i.e., $\mathcal{F}_i = \mathcal{F}_j$ for all $i, j \in
\mathcal{N}$. Finally, we call it a {\em maximum $\sinr$
decentralized feedback policy}, if
$\mathcal{F}_i\paren{\vec{\gamma}_i}$ is only a function of
$\gamma_i^\star$ and $b_i^\star$, and produces a feedback packet
containing $\gamma_i^\star$ as the sole $\sinr$ information on a
positive feedback decision, and otherwise produces $\o$.
\end{definition}

When it is clear from the context, we will omit the term
``general". We will index system-wide feedback policies by
superscripts such as $\vec{\mathcal{F}}^i$, and individual
feedback rules by subscripts such as $\mathcal{F}_i$. We say
$\vec{\mathcal{F}}$ is a {\em beam symmetric} feedback policy if
it satisfies the following condition. Let $\Pi: \R^M \mapsto \R^M$
be a permutation mapping, i.e., $\Pi\paren{\vec{\gamma}} =
\paren{\gamma_{\pi(k)}}_{k=1}^M$ for
some one-to-one $\pi: \brparen{1, 2, \cdots, M} \mapsto \brparen{1, 2, \cdots, M}$.
For $\vec{\Gamma} \in \R^{M \times n}$, let $\Pi\paren{\vec{\Gamma}} = \sqparen{\Pi\paren{\vec{\gamma_1}},
\Pi\paren{\vec{\gamma_2}}, \cdots, \Pi\paren{\vec{\gamma_n}}}$.
If $\mathcal{I}_i$ is the set of beam indexes selected by $\mathcal{F}_i\paren{\vec{\Gamma}}$, then $\pi\paren{\mathcal{I}_i}$
is the set of beam indexes selected by $\mathcal{F}_i\paren{\Pi\paren{\vec{\Gamma}}}$ for all $i \in \mathcal{N}$.
In this paper, we will focus our attention on beam symmetric policies since beams are assumed to be statistically identical.
This is just for the sake of notational simplicity, and the same techniques can be generalized to beam asymmetric policies
by allowing different threshold values for different beams at users.
We let $\Xi$ denote the set of all beam symmetric decentralized feedback policies.
When it is clear from the context, we will also omit the term ``beam symmetric". 
We use the term ``policy" to refer to system-wide feedback rules,
whereas the term ``rule" is used to refer to individual
feedback rules. The definitions given for system-wide feedback
policies extend to individual feedback rules in an obvious way when
possible.

For a feedback policy $\vec{\mathcal{F}}$, we have a random set of users $\mathcal{G}_m\paren{\vec{\mathcal{F}} (\vec{\Gamma})}$ requesting beam $m \in \mathcal{M}$.  Upon feedback, the BS decodes received feedback packets, and selects the best user with the highest $\sinr$ for all $m \in \mathcal{M}$. Therefore, the ergodic downlink sum rate under $\vec{\mathcal{F}}$ is given by
\begin{eqnarray}
R\paren{\vec{\mathcal{F}}}
&=&\EW_{\vec{\Gamma}}\sqparen{r\paren{\vec{\mathcal{F}},\vec{\Gamma}}}
\notag
\\&=& \EW_{\vec{\Gamma}}\sqparen{\sum_{m=1}^M \log\paren{1 + \max_{i \in \mathcal{G}_m\paren{\vec{\mathcal{F}}(\vec{\Gamma})}}
\gamma_{i, m}}} , \label{Eqn: Rate Def}
\end{eqnarray}
where $r\paren{\vec{\mathcal{F}},\vec{\Gamma}}$ is the
instantaneous downlink communication rate under
$\vec{\mathcal{F}}$, and expectation is taken over the random
$\sinr$ matrices. The result of the maximum operation is zero when
$\mathcal{G}_m\paren{\vec{\mathcal{F}}(\vec{\Gamma})}$ is an empty
set, which implies that zero rate is achieved on a particular beam
if no user requests this beam.
$r^{m}\paren{\vec{\mathcal{F}},\vec{\Gamma}}$ and
$R^{m}\paren{\vec{\mathcal{F}}}$ denote the instantaneous
communication rate and the ergodic sum rate on beam $m$,
respectively.  Note that $r^{m}\paren{\vec{\mathcal{F}},
\vec{\Gamma}} = \log\paren{1 + \max_{i \in
\mathcal{G}_m\paren{\vec{\mathcal{F}}(\vec{\Gamma})}} \gamma_{i,
m} }$, and $R^{m}\paren{\vec{\mathcal{F}}} =
\EW_{\vec{\Gamma}}\sqparen{r^{m}\paren{\vec{\mathcal{F}},\vec{\Gamma}}}$.
Also, the sum rate achieved on an event $A$ under
$\vec{\mathcal{F}}$ is written as $ R\paren{\vec{\mathcal{F}}, A}
=
\EW_{\vec{\Gamma}}\sqparen{r\paren{\vec{\mathcal{F}},\vec{\Gamma}}\Inb{A}}$.
We will use $R\paren{\vec{\mathcal{F}}}$ as the performance
measure of a given feedback policy along the rate dimension.

We are interested in maximizing the ergodic sum rate under finite feedback constraints.
To this end, we select the average number of users feeding back per beam $\Lambda(\vec{\mathcal{F}})$
as the performance measure along the feedback dimension.  Since beams are statistically identical,
this measure can be written as $\Lambda(\vec{\mathcal{F}}) = \sum_{i=1}^n \PR{\mathcal{F}_i\paren{\vec{\Gamma}} \mbox{ selects beam } 1}$
for beam symmetric policies.
The resulting rate maximization problem can be written as
\begin{eqnarray}
\begin{array}{ll}
\underset{\vec{\mathcal{F}} \in \Xi}{\mbox{maximize}} & R\paren{\vec{\mathcal{F}}} \\
\mbox{subject to} & \Lambda\paren{\vec{\mathcal{F}}} \leq \lambda
\end{array}. \label{Optimization Problem 1}
\end{eqnarray}
This is not an easy optimization problem to solve since the optimization is over function spaces \cite{Luenberger68}.
Also, the objective function is not necessarily convex \cite{Hazer-WiOpt}.
We will convert the function optimization problem in \eqref{Optimization Problem 1} into an optimization problem over $\R^n$ by proving the optimality of threshold feedback policies.  A general threshold feedback policy is defined as follows.
\begin{definition}\label{thresh policy}
We say $\vec{\mathcal{T}} = \paren{\mathcal{T}_1, \mathcal{T}_2, \cdots, \mathcal{T}_n}^\top$ is a {\em general threshold feedback policy} (GTFP) if, for all $i \in \mathcal{N}$, there is a threshold $\tau_i$ such that $\mathcal{T}_i\paren{\vec{\gamma}_i}$ generates a feedback packet containing $\sinr$ values $\brparen{\gamma_{i, k}}_{k \in \mathcal{I}_i}$ if and only if $\gamma_{i, k} \geq \tau_i$ for all $k \in \mathcal{I}_i \subseteq \mathcal{M}$.  We call it a {\em homogenous general threshold feedback policy} if all users use the same threshold $\tau$, i.e., $\tau_i =\tau$ for all $i \in \mathcal{N}$.
\end{definition}


We note that a user can be allocated to multiple beams according to this definition. Another class of threshold feedback policies limiting each user to request at most one beam, which is, of course, the beam achieving the maximum $\sinr$, has been widely studied in the literature \cite{hassibi, Pugh10, Diaz06, Tharaka-ICC}.  We call this class of feedback policies maximum $\sinr$ threshold feedback policies, and formally define them as follows.
\begin{definition}\label{max sinr thresh policy}
$\vec{\mathcal{T}} = \paren{\mathcal{T}_1, \mathcal{T}_2, \cdots, \mathcal{T}_n}^\top$ is a {\em maximum $\sinr$ threshold feedback policy} (MTFP) if, for all $i \in \mathcal{N}$, there is a threshold $\tau_i$ such that $\mathcal{T}_i\paren{\vec{\gamma}_i}$ produces a feedback packet requesting beam $m$ and containing $\gamma_{i, m}$ as the sole $\sinr$ information if and only if $b_i^\star = m$ and $\gamma_i^\star \geq \tau_i$.
\end{definition}

For a given set of threshold values, it is not hard to see that the GTFP (corresponding to these threshold values) always achieves a rate at least as good as the rate achieved by the MTFP (corresponding to the same threshold values)  because users request all the beams with $\sinr$ values above their thresholds under GTFP, which includes the best beam with the highest $\sinr$.  Since maximum $\sinr$ values are also fed back by GTFPs, they can be considered more general than MTFPs.  Moreover, it can also be shown that a GTFP reduces to an MTFP if threshold values of all users are greater than one.  In Section \ref{Section: Optimality of Thresh}, we will prove that the set of all GTFPs and the set of all MTFPs are rate-wise optimal subsets of general decentralized feedback policies, and maximum $\sinr$ decentralized feedback policies, respectively.

\section{Optimality of Threshold Feedback Policies}\label{Section: Optimality of Thresh}

\subsection{General Threshold Feedback Policies}

We will focus on the first beam to explain our proof ideas without any loss of generality since all beams are statistically identical, and feedback policies are beam symmetric.  
For our proofs, we will define various sets whose elements lie in various spaces including $\Rp^M$ and $\Rp^{M \times n}$.  Therefore, paying attention to in which space the elements of a set lie will facilitate the exposition in the rest of the paper.

For a given beam symmetric general decentralized feedback policy $\vec{\mathcal{F}} = \paren{\mathcal{F}_1, \mathcal{F}_2, \cdots, \mathcal{F}_n}^\top$, we let $FB_i=\brparen{\vec{\gamma}_i \in \Rp^{M}:  \mathcal{F}_i\paren{\vec{\gamma}_i} \mbox{ selects beam } 1}$ for all $i \in \mathcal{N}$.  Given $\vec{\mathcal{F}}$, we construct a GTFP $\vec{\mathcal{T}}$ by choosing $\tau_i$ as $\PR{\gamma_{i,1}\geq \tau_i}=\PR{\vec{\gamma}_{i} \in FB_i}$ for all $i \in \mathcal{N}$.  This construction is feasible since $\gamma_{i, 1}$ has a continuous distribution function.  Such a selection of $\vec{\mathcal{T}}$ leads to a fair comparison between $\vec{\mathcal{F}}$ and $\vec{\mathcal{T}}$ since $\Lambda(\vec{\mathcal{F}})=\Lambda(\vec{\mathcal{T}})$.  We divide $FB_i$ into two disjoint sets $\mathcal{S}_i^L = \brparen{\vec{\gamma}_i  \in \Rp^M: \vec{\gamma}_i \in FB_i \ \&\ \gamma_{i, 1} <\tau_i}$, and $\mathcal{S}_i^R = \brparen{\vec{\gamma}_i \in \Rp^M: \vec{\gamma}_i \in FB_i \ \&\ \gamma_{i, 1} \geq\tau_i}$.
Finally, we let $\bar{\mathcal{S}}_i^R = \brparen{\vec{\gamma}_i \in \Rp^M: \vec{\gamma}_i \notin FB_i \ \&\ \gamma_{i, 1} \geq\tau_i}$. Using these sets, we will show that $R\paren{\vec{\mathcal{T}}} \geq R\paren{\vec{\mathcal{F}}}$.

The proof of this result for a simple single user single beam system is easy.  For a particular realization of the $\sinr$ value $\gamma_1$, the system, or the user in this case, will achieve the same instantaneous rate through both feedback policies if they result in the same feedback decision, i.e., when both policies lead to a positive or negative feedback decision together for this $\sinr$ value.  On the other hand, the achieved instantaneous rate will be clearly different if only one of the policies results in a positive feedback decision.  This happens either when $\gamma_1 \in \mathcal{S}^L_1$, in which case only $\vec{\mathcal{F}}$ leads to a positive feedback decision, or when $\gamma_1 \in \bar{\mathcal{S}}_1^R$, in which case only $\vec{\mathcal{T}}$ leads to a positive feedback decision.  The worst case $\sinr$ achieved by the user on the event $\gamma_1 \in \bar{\mathcal{S}}_1^R$ is greater than the threshold value $\tau_1$, and the best case $\sinr$ achieved by the user on the event $\gamma_1 \in \mathcal{S}^L_1$ is less than $\tau_1$.  Therefore, the rates achieved by $\vec{\mathcal{F}}$ and $\vec{\mathcal{T}}$ can be upper and lower bounded, respectively, to show that $R\paren{ \vec{\mathcal{T}} } \geq R\paren{ \vec{\mathcal{F}} }$. 

The proof for the multiuser scenario hinges on the same principles above but it is not straightforward due to coupling effects of individual feedback policies on the aggregate rate expression.  Part of the complexity to deal with these effects arises from the non-homogenous nature of the feedback rules.  
To overcome this problem, we will prove a more general result indicating that the best strategy for user $i$ is to always use a threshold feedback rule whatever the feedback rules of other users are.

To this end, for a given $\vec{\mathcal{F}} =\paren{\mathcal{F}_1,\mathcal{F}_2,\cdots,\mathcal{F}_n}^\top$, we let $\mathcal{G}_1^{-1}\paren{\vec{\mathcal{F}}(\vec{\Gamma})}=\brparen{i \in \mathcal{N}: i \neq 1\ \&\  i \in \mathcal{G}_1\paren{\vec{\mathcal{F}}(\vec{\Gamma})}}$.  That is, $\mathcal{G}_1^{-1}\paren{\vec{\mathcal{F}}(\vec{\Gamma})}$ is the random set of users containing all users requesting beam $1$ under $\vec{\mathcal{F}}$, except for the first user. The superscript $-1$ is used to indicate that all users but user $1$ requesting beam $1$ are included in $\mathcal{G}_1^{-1}\paren{\vec{\mathcal{F}}(\vec{\Gamma})}$.  The maximum beam $1$ $\sinr$ value achieved by a user in this random set is denoted by $\bar{\gamma}_1^\star\paren{\vec{\mathcal{F}}}$, and it is equal to $\bar{\gamma}_1^\star\paren{\vec{\mathcal{F}}}= \max_{i \in \mathcal{G}_1^{-1}\paren{\vec{\mathcal{F}}(\vec{\Gamma})}} \gamma_{i, 1}$.

Consider now another decentralized feedback policy $\vec{\mathcal{F}}^1=\paren{\mathcal{T}_1,\mathcal{F}_2,\cdots,\mathcal{F}_n}^\top$.  That is, we only allow user $1$ to switch to the threshold feedback rule $\mathcal{T}_1$ with threshold value $\tau_1$ determined as above. Then, for almost all realizations of $\vec{\Gamma}$, we have $\bar{\gamma}_1^\star\paren{\vec{\mathcal{F}}} =\bar{\gamma}_1^\star\paren{\vec{\mathcal{F}}^1}=\bar{\gamma}_1^\star$. 
Therefore, the difference between $R\paren{\vec{\mathcal{F}}}$ and $R\paren{\vec{\mathcal{F}}^1}$ depends only on the rate achieved by user $1$ under two feedback policies.

We are interested in proving $R\paren{\vec{\mathcal{F}}} \leq R\paren{\vec{\mathcal{T}}}$.  A brief sketch of the proof is as follows.  We will first prove that $R\paren{\vec{\mathcal{F}}} \leq R\paren{\vec{\mathcal{F}}^1}$.  Let $\vec{\Gamma}_{-1}$ be the $\sinr$ matrix containing $\sinr$ values of all users except those of the first user.  Let also $R\paren{\vec{\mathcal{F}} | \vec{\Gamma}_{-1}} = \EW_{\vec{\Gamma}}\sqparen{r\paren{\vec{\mathcal{F}},\vec{\Gamma}} | \vec{\Gamma}_{-1}}$ be the conditional average rate achieved by $\vec{\mathcal{F}}$ for a given $\vec{\Gamma}_{-1}$.  Then, it is enough to show that $R\paren{\vec{\mathcal{F}}^1 | \vec{\Gamma}_{-1}} \geq R\paren{\vec{\mathcal{F}} | \vec{\Gamma}_{-1}}$ for almost all $\vec{\Gamma}_{-1}$.  This result implies that the aggregate communication rate increases if user $1$ switches to a threshold feedback rule regardless of feedback rules of other users.  Repeating the same steps for other users $i \in \brparen{2, 3, \cdots, n}$ one-by-one, we end up with the threshold feedback policy $\vec{\mathcal{T}}$ after $n$ steps, and conclude that $R\paren{\vec{\mathcal{T}}} \geq R\paren{\vec{\mathcal{F}}}$.

Before giving the proof details, we will first perform a preliminary analysis.  For the rest of this part of the paper, $\vec{\mathcal{F}}^1$ will represent the decentralized feedback policy derived from a given decentralized feedback policy $\vec{\mathcal{F}}$ as above.  When we switch from $\vec{\mathcal{F}}$ to $\vec{\mathcal{F}}^1$, we can identify three main types of events: {\em neutral}, {\em loss} and {\em gain} events.  On the neutral event, we will continue to achieve the same downlink throughput under both feedback policies.  On the loss event, we will lose some data rate when we switch from $\vec{\mathcal{F}}$ to $\vec{\mathcal{F}}^1$.  Finally, on the gain event,  we will gain some data rate when we switch from $\vec{\mathcal{F}}$ to $\vec{\mathcal{F}}^1$.  The difference $R\paren{\vec{\mathcal{F}}^1} - R\paren{\vec{\mathcal{F}}}$ depends on the average rates lost and gained on the loss and gain events.  To show that $R\paren{\vec{\mathcal{F}}^1} - R\paren{\vec{\mathcal{F}}} \geq 0$, we need to characterize loss and gain events precisely.  We first formally define these events, and then provide their further characterizations suitable for our analysis in Lemmas \ref{loss event lemma_MB} and \ref{gain event lemma_MB}.

\begin{definition}\label{events_MB}
When we switch from $\vec{\mathcal{F}}$ to $\vec{\mathcal{F}}^1$, the loss, gain and neutral events on beam $1$ are defined as
\begin{eqnarray}
 A_L= \brparen{\vec{\Gamma} \in \Rp^{M \times n}\  :\ r^1\paren{\vec{\mathcal{F}}^1, \vec{\Gamma}} < r^1\paren{\vec{\mathcal{F}},\vec{\Gamma}}}, \label{loss event_MB} \\
A_G= \brparen{\vec{\Gamma} \in \Rp^{M \times n}\   : \ r^1\paren{\vec{\mathcal{F}}^1, \vec{\Gamma}} > r^1\paren{\vec{\mathcal{F}},\vec{\Gamma}}} \ \ \label{gain event_MB}
\end{eqnarray}
and
\begin{eqnarray}
A_N= \brparen{\vec{\Gamma} \in \Rp^{M \times n}\  :\ r^1(\vec{\mathcal{F}}^1,\vec{\Gamma}) = r^1(\vec{\mathcal{F}},\vec{\Gamma})},\label{neutral event_MB}
\end{eqnarray}
respectively.
\end{definition}


The next two lemmas provide other characterizations of the loss and gain events.  These characterizations will be important when we compare $R\paren{\vec{\mathcal{F}}^1}$ against $R\paren{\vec{\mathcal{F}}}$.  We skip their proofs due to space limitations.
\begin{lemma} \label{loss event lemma_MB}
$A_L$ is equal to
\begin{eqnarray*}
A_L = \brparen{\vec{\Gamma} \in \Rp^{M \times n} \  : \  \vec{\gamma}_{1} \in \mathcal{S}_1^L\quad \& \quad \bar{\gamma}_1^\star < \gamma_{1,1}}.
\end{eqnarray*}
\end{lemma}
\begin{lemma} \label{gain event lemma_MB}
$A_G$ is equal to
\begin{eqnarray*}
A_G = \brparen{\vec{\Gamma} \in \Rp^{M \times n} \  : \  \vec{\gamma}_{1} \in \bar{\mathcal{S}}_1^R \quad\& \quad \bar{\gamma}_1^\star < \gamma_{1,1}}.
\end{eqnarray*}
\end{lemma}

These auxiliary results will help us to prove the optimality of $\vec{\mathcal{F}}^1$ over $\vec{\mathcal{F}}$ in terms of sum rate in Theorem \ref{Optimality of Thresholding Rules- NonHomo}.  Before providing the details of the proof of this theorem, we will again give a sketch of the proof.  $A_L$, $A_G$ and $A_N$ are three disjoint events with total probability mass of one. Therefore, for a feedback policy $\vec{\mathcal{F}}$, we can write $R^1\paren{\vec{\mathcal{F}} | \vec{\Gamma}_{-1}}  = R^1\paren{\vec{\mathcal{F}}, A_L | \vec{\Gamma}_{-1}} + R^1\paren{\vec{\mathcal{F}}, A_G | \vec{\Gamma}_{-1}} + R^1\paren{\vec{\mathcal{F}}, A_N | \vec{\Gamma}_{-1}}$.
We can write a similar expression for $R^1\paren{\vec{\mathcal{F}}^1 | \vec{\Gamma}_{-1}}$, and the comparison of these two equations term-by-term reveals that $R^1\paren{\vec{\mathcal{F}}^1 | \vec{\Gamma}_{-1}} \geq R^1\paren{\vec{\mathcal{F}} | \vec{\Gamma}_{-1}}$.  Since this inequality holds for almost all $\vec{\Gamma}_{-1}$, we also have $R^1\paren{\vec{\mathcal{F}}^1} \geq R^1\paren{\vec{\mathcal{F}}}$.  Since beams are identical, the total rate is $M$ times the rate achieved on beam $1$.  Therefore, we finally have $R\paren{\vec{\mathcal{F}}^1} \geq R\paren{\vec{\mathcal{F}}}$.  We make this idea formal in the proof of the next theorem.

\begin{theorem}
\label{Optimality of Thresholding Rules- NonHomo}
Let $\vec{\mathcal{F}} = \paren{\mathcal{F}_1,\mathcal{F}_2,\cdots,\mathcal{F}_n}^\top$ and $\vec{\mathcal{F}}^1=\paren{\mathcal{T}_1,\mathcal{F}_2,\cdots,\mathcal{F}_n}^\top$ be defined as above.  Then, $\Lambda\paren{\vec{\mathcal{F}}} = \Lambda\paren{\vec{\mathcal{F}}^1}$, and $R\paren{\vec{\mathcal{F}}^1} \geq R\paren{\vec{\mathcal{F}}}$ for any $M \geq 1$.
\end{theorem}
\begin{IEEEproof}
It is enough to prove $R^1\paren{\vec{\mathcal{F}}^1 | \vec{\Gamma}_{-1}} \geq R^1\paren{\vec{\mathcal{F}} | \vec{\Gamma}_{-1}}$ for almost all $\vec{\Gamma}_{-1}$.  By definition, we have $R^1\paren{\vec{\mathcal{F}}, A_N |\vec{\Gamma}_{-1}} = R^1\paren{\vec{\mathcal{F}}^1, A_N | \vec{\Gamma}_{-1}}$, and therefore we are only interested on the average rates on loss and gain events.

The following identity follows from the definition of conditional expectation.
\begin{eqnarray*}
R^1\paren{\vec{\mathcal{F}}, A_L | \vec{\Gamma}_{-1}} =\PRW\paren{A_L | \vec{\Gamma}_{-1}}\EW_{\vec{\Gamma}}\sqparen{r^1\paren{\vec{\mathcal{F}},\vec{\Gamma}} | A_L, \vec{\Gamma}_{-1}}.
\end{eqnarray*}
Lemma \ref{loss event lemma_MB} implies that whenever $A_L$ is correct, the user $1$ requests beam $1$, and achieves the best $\sinr$ on beam $1$ among all the users requesting beam $1$.  Since $\vec{\gamma}_{1} \in \mathcal{S}_1^L$ on $A_L$, $\gamma_{1, 1}$ is less than $\tau_1$. Therefore,
\begin{eqnarray}
R^1\paren{\vec{\mathcal{F}}, A_L | \vec{\Gamma}_{-1}} &\leq& \PRW\paren{A_L|\vec{\Gamma}_{-1}} \log{\paren{1 + \tau_1}}. \label{Rn_AL_F}
\end{eqnarray}
We can similarly write
\begin{eqnarray*}
R^1\paren{\vec{\mathcal{F}}, A_G | \vec{\Gamma}_{-1}} = \PRW\paren{A_G | \vec{\Gamma}_{-1}}\EW_{\vec{\Gamma}}\sqparen{r^1\paren{\vec{\mathcal{F}},\vec{\Gamma}} | A_G, \vec{\Gamma}_{-1}}.
\end{eqnarray*}
Lemma \ref{gain event lemma_MB} implies that user $1$ achieves the best $\sinr$ on beam $1$ among all the users requesting beam $1$ but $\vec{\gamma}_{1} \in \bar{\mathcal{S}}_1^R$ on $A_G$. Therefore, $\vec{\gamma}_{1} \notin FB_1$, and user $1$ will not request beam $1$ under $\vec{\mathcal{F}}$.  Hence, $\vec{\mathcal{F}}$ schedules beam $1$ to the user with $\sinr$ value $\bar{\gamma}_1^\star$, which leads to \footnote{Note that $\bar{\gamma}_1^\star$ is a (measurable) function of $\vec{\Gamma}_{-1}$, and therefore \eqref{Rn_AG_F} conforms with the measure theoretic definition of the conditional expectation.}
\begin{eqnarray} R^1\paren{\vec{\mathcal{F}}, A_G | \vec{\Gamma}_{-1}} &=& \PRW\paren{A_G | \vec{\Gamma}_{-1}} \log{\paren{1 + \bar{\gamma}_1^\star}}. \label{Rn_AG_F}
\end{eqnarray}

Similar to the above arguments, user $1$ will not request beam $1$ under $\vec{\mathcal{F}}^1$ on the event $A_L$ since $\vec{\gamma}_{1} \in \mathcal{S}_1^L$.  This means
\begin{eqnarray}
R^1\paren{\vec{\mathcal{F}}^1, A_L | \vec{\Gamma}_{-1}}
= \PRW\paren{A_L|\vec{\Gamma}_{-1}} \log{\paren{1 + \bar{\gamma}_1^\star}}.\label{Rn_AL_F1}
\end{eqnarray}
Finally, user $1$ requests beam $1$ under $\vec{\mathcal{F}}^1$ on $A_G$, leading to
\begin{eqnarray}
R^1\paren{\vec{\mathcal{F}}^1, A_G | \vec{\Gamma}_{-1}}
\geq \PRW\paren{A_G|\vec{\Gamma}_{-1}} \log{\paren{1 + \max\paren{\tau_1,\bar{\gamma}_1^\star}}}. \label{Rn_AG_F1}
\end{eqnarray}
The purpose of the maximum operator is to obtain a tighter lower bound.  By using \eqref{Rn_AL_F}, \eqref{Rn_AG_F}, \eqref{Rn_AL_F1} and \eqref{Rn_AG_F1}, we can write
\begin{eqnarray}
\lefteqn{R^1\paren{\vec{\mathcal{F}}^1 | \vec{\Gamma}_{-1}} - R^1\paren{\vec{\mathcal{F}} | \vec{\Gamma}_{-1}} \geq} \hspace{8cm} \nonumber \\
\lefteqn{\PRW\paren{A_G | \vec{\Gamma}_{-1}}\paren{\log{\paren{1 + \max\paren{\tau_1,\gamma_1^\star}}} - \log{\paren{1 + \gamma_1^\star}}}} \hspace{7.5cm} \nonumber \\
\lefteqn{+\PRW\paren{A_L|\vec{\Gamma}_{-1}}\paren{\log{\paren{1 + \bar{\gamma}_1^\star}}-\log{\paren{1 + \tau_1}}}.} \hspace{7.8cm} \nonumber
\end{eqnarray}

To conclude the proof, we need to analyze two different cases separately.  If $\bar{\gamma}_1^\star \geq \tau_1$, then it directly follows that $R^1\paren{\vec{\mathcal{F}}^1 | \vec{\Gamma}_{-1}}-R^1\paren{\vec{\mathcal{F}} | \vec{\Gamma}_{-1}} \geq 0$.  If $\bar{\gamma}_1^\star < \tau_1$, then we have
\begin{eqnarray}
\lefteqn{R^1\paren{\vec{\mathcal{F}}^1 | \vec{\Gamma}_{-1}} - R^1\paren{\vec{\mathcal{F}} | \vec{\Gamma}_{-1}} \geq} \hspace{8cm} \nonumber \\
\lefteqn{\paren{\PRW\paren{A_G|\vec{\Gamma}_{-1}} - \PRW\paren{A_L|\vec{\Gamma}_{-1}}}\paren{\log\paren{1+\tau_1} - \log\paren{1+\gamma_1^\star}}} \hspace{7.8cm} \nonumber
\end{eqnarray}
Since $\bar{\gamma}_1^\star < \tau_1$, we have $\PRW\paren{A_G | \vec{\Gamma}_{-1}} = \PR{\vec{\gamma}_{1} \in
\bar{\mathcal{S}}_1^R}$, and $\PRW\paren{A_L | \vec{\Gamma}_{-1}} \leq \PR{\vec{\gamma}_{1}\in \mathcal{S}_1^L}$.  Since $
\PR{\vec{\gamma}_{1}\in \bar{\mathcal{S}}_1^R}=\PR{\vec{\gamma}_{1}\in \mathcal{S}_1^L}$, we also have $R^1\paren{\vec{\mathcal{F}}^1 | \vec{\Gamma}_{-1}} - R^1\paren{\vec{\mathcal{F}} | \vec{\Gamma}_{-1}} \geq 0$ for $\bar{\gamma}_1^\star < \tau_1$.
This proves that $R^1\paren{\vec{\mathcal{F}}^1 | \vec{\Gamma}_{-1}} \geq R^1\paren{\vec{\mathcal{F}} | \vec{\Gamma}_{-1}}$ for almost all $\vec{\Gamma}_{-1}$. 
\end{IEEEproof}

This theorem shows that if a user starts using a threshold feedback rule, the system will benefit in terms of ergodic sum rate regardless of the feedback rules all other users.  This result leads to the following key finding.

\begin{theorem} \label{Thm: Optimality of Thresholding Rules}
For any beam symmetric general decentralized feedback policy $\vec{\mathcal{F}}$, there exists a GTFP $\vec{\mathcal{T}}$ such that $\Lambda\paren{\vec{\mathcal{F}}} = \Lambda\paren{\vec{\mathcal{T}}}$ and $R\paren{\vec{\mathcal{T}}} \geq R\paren{\vec{\mathcal{F}}}$.
\end{theorem}
\begin{IEEEproof}
For a given $\vec{\mathcal{F}} = \paren{\mathcal{F}_1, \mathcal{F}_2, \cdots, \mathcal{F}_n}^\top$, let $\vec{\mathcal{T}} = \paren{\mathcal{T}_1, \mathcal{T}_2, \cdots, \mathcal{T}_n}^\top$ be the GTFP constructed as above.  Let $\vec{\mathcal{F}}^{k} = \paren{\mathcal{T}_1, \cdots, \mathcal{T}_k, \mathcal{F}_{k+1}, \cdots, \mathcal{F}_n}^\top$ for $1 \leq k \leq n$.  When $k = n$, we have $\vec{\mathcal{F}}^n = \vec{\mathcal{T}}$.  By Theorem \ref{Optimality of Thresholding Rules- NonHomo}, we have $R\paren{\vec{\mathcal{F}}} \leq R\paren{\vec{\mathcal{F}}^1} \leq \cdots \leq R\paren{\vec{\mathcal{F}}^n}$.  Since $\Lambda\paren{\vec{\mathcal{F}}} = \Lambda\paren{\vec{\mathcal{F}}^1} = \cdots = \Lambda\paren{\vec{\mathcal{F}}^n}$, the proof is complete.
\end{IEEEproof}

\subsection{Maximum $\sinr$ Threshold Feedback Policies}

In this part, we will provide similar results for MTFPs.  Under a maximum $\sinr$ decentralized feedback policy, each user requests only the beam achieving the maximum $\sinr$ if the feedback conditions are met, e.g., see Definitions \ref{fb policy} and \ref{max sinr thresh policy}. Proving the optimality of MTFPs in the set of maximum $\sinr$ feedback policies is similar to the proof we gave for the optimality of GTFPs in Theorems \ref{Optimality of Thresholding Rules- NonHomo} and \ref{Thm: Optimality of Thresholding Rules}.  There are only some subtle differences. Now, the thresholds are set such that $\PR{b_i^\star = 1 \mbox{ and } \gamma_i^\star \geq \tau_i}=\PR{\vec{\gamma_{i}} \in FB_i}$.  The definition of $FB_i$ can be further refined in which user $i$ requests beam $1$ if and only if $b_i^\star = 1$ and $\gamma_i^\star$ satisfies some feedback conditions.  The definitions of other sets and events of interest require only some subtle modifications, too.  For example, $A_L$ can now be defined as $A_L = \brparen{\vec{\Gamma} \in \Rp^{M \times n} \  : \  \vec{\gamma}_{1} \in \mathcal{S}_1^L\quad \& \quad \bar{\gamma}_1^\star < \gamma_1^\star}$ where $\mathcal{S}_1^L = \brparen{\vec{\gamma}_1 \in \Rp^M: \vec{\gamma}_1 \in FB_1 \ \&\ \gamma_1^\star <\tau_1}$. The next two theorems provide results analogous to the ones stated in Theorems \ref{Optimality of Thresholding Rules- NonHomo} and \ref{Thm: Optimality of Thresholding Rules}.

\begin{theorem}
\label{Optimality of Thresholding Rules- NonHomo_MTFP}
For a given beam symmetric decentralized maximum $\sinr$ threshold feedback policy $\vec{\mathcal{F}} = \paren{\mathcal{F}_1,\mathcal{F}_2,\cdots,\mathcal{F}_n}^\top$, let  $\vec{\mathcal{F}}^1 = \paren{\mathcal{T}_1,\mathcal{F}_2,\cdots,\mathcal{F}_n}^\top$ be another maximum $\sinr$ threshold feedback policy derived from $\vec{\mathcal{F}}$ by allowing user $1$ to switch from $\mathcal{F}_1$ to $\mathcal{T}_1$, where $\mathcal{T}_1$ is a beam symmetric maximum $\sinr$ threshold rule whose threshold is set as above.  Then, $\Lambda\paren{\vec{\mathcal{F}}} = \Lambda\paren{\vec{\mathcal{F}}^1}$, and $R\paren{\vec{\mathcal{F}}^1} \geq R\paren{\vec{\mathcal{F}}}$ for any $M \geq 1$.
\end{theorem}

\begin{theorem} \label{Thm: Optimality of MTFP Thresholding Rules}
For any beam symmetric decentralized maximum $\sinr$ feedback policy $\vec{\mathcal{F}}$, there exists an MTFP $\vec{\mathcal{T}}$ such that $\Lambda\paren{\vec{\mathcal{F}}} = \Lambda\paren{\vec{\mathcal{T}}}$ and $R\paren{\vec{\mathcal{T}}} \geq R\paren{\vec{\mathcal{F}}}$.
\end{theorem}

\section{Discussion of Results}\label{Section: Discussion}
In this part, we will briefly discuss the results presented in Section \ref{Section: Optimality of Thresh}.  We start with a comparison between GTFPs and MTFPs.  The main advantage of GTFPs over MTFPs is the ability of the BS to allocate multiple beams to a user.  Therefore, a GTFP policy achieves higher data rates when compared to an MTFP policy with the same threshold levels.  From a practical point of view, such gains in data rates are expected to be minor due to dependencies among beams at a user, i.e., high $\gamma_{i, m}$ implies low $\gamma_{i, k}$.  Moreover, both types of policies achieve the same performance if all threshold values are larger than $1$ ($0$dB), which is a realistic figure in a practical system.

From a theoretical point of view, the resulting optimization problem over $\Rp^n$ lends itself more amenable to further analysis if we focus on GTFPs.  More specifically, we can search for the optimal beam symmetric feedback policies within the class of GTFPs without sacrificing from optimality thanks to Theorem \ref{Thm: Optimality of Thresholding Rules}, and with a slight abuse of notation, we can equivalently write \eqref{Optimization Problem 1} as
\begin{eqnarray}
\begin{array}{ll}
\underset{\vec{\tau} \in \Rp^n}{\mbox{maximize}} & R\paren{\vec{\tau}} \\
\mbox{subject to} & \sum_{i = 1}^n \PR{\gamma_{i, 1} \geq \tau_i}
\leq \lambda
\end{array}. \label{Optimization Problem 2}
\end{eqnarray}


In \cite{Tharaka-ISIT2}, we show that the rate becomes a
Schur-concave function of feedback probabilities $p_i =
\PR{\gamma_{i, 1} \geq \tau_i}$ if the $\sinr$ distribution
satisfies some mild conditions, and therefore establish the
optimality of homogenous general threshold feedback policies among
the class of beam symmetric general decentralized feedback
policies.  In general, a homogenous threshold feedback policy,
with some surprise, is not always a solution for
\eqref{Optimization Problem 2} even if all users experience
statistically identical channel conditions. However, for Rayleigh
fading channels with $M \geq 2$, the solution for
\eqref{Optimization Problem 2} is always a homogenous threshold
feedback policy.  Interested readers are referred to
\cite{Tharaka-ISIT2} for further details on these results.

A key feature of our results in this paper is that they are distribution independent, and hold for most practical fading distributions such as Rayleigh, Ricean and Nakagami distributions.

Some further game theoretic insights are as follows.  We will only focus on GTFPs but similar explanations also hold for MTFPs.  Given the same utility function $R\paren{\mathcal{F}_1, \mathcal{F}_2, \cdots, \mathcal{F}_n}$ for all users, the selfish optimization problem faced by user $i$ can be stated as choosing a beam symmetric decentralized feedback rule maximizing her utility given other users' feedback rules without increasing the feedback level.  Theorem \ref{Optimality of Thresholding Rules- NonHomo} shows that the dominant strategy is to switch from $\mathcal{F}_i$ to the corresponding threshold rule $\mathcal{T}_i$.  As a result, the set of GTFPs constitute the set of Nash equilibria for this feedback rule selection game, and therefore GTFPs are also stable operating points from a game theoretic point of view.

\section{Conclusions}\label{Section: Conclusions}
In this paper, we have established the structure of rate-wise optimal decentralized feedback policies for opportunistic vector broadcast channels under finite feedback constraints.  In particular, we have shown that threshold feedback policies are optimal to maximize data rates for such channels under finite feedback constraints.  Our results do not depend on specific fading statistics, and hold for most common fading distributions including Rayleigh, Ricean and Nakagami distributions.  They form an analytical justification for the use of threshold feedback policies in practical systems, and reinforce previous work on thresholding as a selective feedback policy to reduce feedback levels. Since each threshold feedback policy can be associated with a threshold vector in $\Rp^n$, these results also reduce the search for rate-wise optimal feedback policies from function spaces to finite dimensional Euclidean spaces.

\bibliography{Tharaka-bibfile}
\end{document}